\documentclass[11pt]{article}

\usepackage{amsmath,amssymb,exscale}
\usepackage{marvosym}
\usepackage{array,multicol}
\usepackage{afterpage,float,flafter}
\usepackage{amssymb}
\usepackage{epsfig,rotating,pifont}
\usepackage{cite}
\usepackage{hyperref}
\usepackage{color}
\usepackage{ marvosym }
\usepackage{textcomp}
\usepackage{wrapfig}

\definecolor{nicered}{rgb}{0.5,0.1,0.1}
\definecolor{nicegreen}{rgb}{0.1,0.5,0.1}
\definecolor{niceblue}{rgb}{0.1,0.1,0.8}
\hypersetup{colorlinks,citecolor= nicegreen,linkcolor= nicered}

\@addtoreset{equation}{section} \makeatother

\def\beq{\begin{eqnarray}}
\def\eeq{\end{eqnarray}}
\def\lsim{\mathrel{\rlap{\lower3pt\hbox{\hskip0pt$\sim$}}
\raise1pt\hbox{$<$}}}         
\def\gsim{\mathrel{\rlap{\lower4pt\hbox{\hskip1pt$\sim$}}
\raise1pt\hbox{$>$}}}         

\title{
\vspace{-3cm}
\begin{flushright}
\small{CERN-PH-TH/2012-067}
\end{flushright}
\vspace{2.7cm}
\begin{center}
\medskip
{\Huge\bf Composite MFV and Beyond}
\end{center}
\vspace{0.6cm}
\author{
\Large{\text{\bf Michele Redi$^{1,2}$}\footnote{michele.redi@cern.ch}~~
}
\\ \\
$^1$\emph{CERN, Theory Division, CH-1211, Geneva 23, Switzerland}\\
$^2$\emph{INFN, 50019 Sesto F., Firenze, Italy}\\
}
}

\date{}
\begin{document}
\maketitle \thispagestyle{empty} \vspace*{-.2cm}

\begin{abstract}
We revisit and extend realizations of Minimal Flavor Violation (MFV) in theories with strongly
coupled electro-weak symmetry breaking. MFV requires that some chiralities of 
light SM quarks are strongly composite leading, depending on the scenario, to bounds from 
compositeness searches, precision electro-weak tests or even flavor physics.
Within the framework of partial compositeness we show how to extend the MFV paradigm  
allowing the treat the top quark differently. This can be  realized if for example the strong 
sector has an $U(2)$ symmetry. In this case the light generations can be mostly elementary and
all the bounds are easily satisfied.
\end{abstract}

\newpage
\renewcommand{\thepage}{\arabic{page}}
\setcounter{page}{1}

\section{Introduction}

Arguably flavor physics is the main outstanding problem of all extensions of the Standard Model (SM) that solve the 
hierarchy problem. This is especially true for strongly coupled theories of electro-weak symmetry breaking, 
that up to recently were considered at odds with flavor data.

The idea of partial compositeness greatly improves this situation \cite{flavorgeneration,minimalcomposite}. 
In these scenarios (see \cite{RSreview} for reviews), the SM fields are partially composite due to mixings with the states of
the composite sector that is ultimately responsible for the breaking of the electro-weak symmetry of the SM.
If the strong sector is anarchic, i.e. it has no hierarchies, the hierarchical SM Yukawa couplings must originate 
from the mixings so that the light generations are naturally mostly elementary, while the third 
generation is significantly or even entirely composite. The small amount of compositeness of the light 
quarks suppresses flavor violations of the first two generations, roughly in agreement with  
data despite the fact that many more flavor structures exist than in the SM. 
Bounds from precision measurements of SM couplings can also be satisfied for the same reason.  

In the anarchic scenario, flavor bounds are generically respected with few tensions, in particular CP violation
in the Kaon system, Electric Dipole Moments (EDMs), and lepton flavor violation, see \cite{weiler}.
We cannot at present quantify how worrisome these tensions are but if the paradigm of anarchic partial compositeness 
is connected to the electro-weak scale, as demanded  by naturalness, new flavor physics must be around 
the corner in various observables. Another possibility is that a different flavor structure 
is realized in these theories. As shown in Refs. \cite{compositeMFV,Cacciapaglia:2007fw} partial compositeness also allows attractive realizations 
of the Minimal Flavor Violation (MFV) \cite{MFV}. If the strong sector respects appropriate flavor symmetries, the only flavor structures 
are associated to the mixings and MFV can be realized if only two mixings break the SM flavor symmetry. 
Another possibility considered in the literature is that the SM quarks are all composite and the flavor symmetries are 
broken similarly to the SM \cite{rattazzizaffaroni}.

Differently from the anarchic scenarios, with MFV at least some of the light quarks have large compositeness, 
being related by the flavor symmetry  to the one of the top.  As a consequence composite MFV is typically strongly constrained 
by precision electro-weak measurements. One notable exception is the case of right-handed compositeness proposed 
in \cite{compositeMFV}, where large corrections to the SM couplings are not generated and milder bounds currently 
follow from recent searches at the LHC. In this note we introduce another category of theories where we treat the top quark differently, allowing 
the light generations to be mostly elementary compatibly with the flavor symmetry, see \cite{barbieri} for a realization in supersymmetry.
These theories are based on a flavor symmetry which is a subgroup of the SM flavor symmetries, for example 
$U(3)_L\otimes U(3)_{Rd}\otimes U(2)_{Ru}$. As we will see flavor bounds are similar to MFV but,
being the first two generations mostly elementary, all limits from direct searches and precision tests
are easily satisfied. These scenarios will be more challenging to test at the LHC with a phenomenology
similar to the anarchic models.

The paper is organized as follows. In section \ref{pcflavor} we discuss the general properties of flavor in 
theories with partial compositeness. In section \ref{compositeMFV} we review known realizations of MFV
in these scenarios and present a general analysis of the effective operators relevant for flavor physics. 
We study extensions of MFV which allow small compositeness of light quarks in section \ref{beyondMFV}, showing that
they can be in agreement with flavor, EDMs and precision tests. We conclude in \ref{outlook}. 

\section{Flavor and Partial Compositeness}
\label{pcflavor}

The framework of partial compositeness allows new realizations of the SM flavor structure in strongly 
coupled theories of electro-weak symmetry breaking. This is a major improvement compared to previous constructions 
that allows these theories to be in rough agreement with the stringent flavor bounds. We here review the basic ingredients 
focusing  on the relevant symmetries\footnote{We will use a purely four-dimensional language throughout this paper, 
keeping in mind that all the scenarios described here  can be realized as theories in 5D (see appendix in \cite{compositeMFV}).}.

Our discussion applies to theories with a composite Higgs, Higgsless, or extended Higgs sectors \cite{cthdm}. 
The flavor structure of models with partial compositeness can be understood in terms of symmetries similarly to the SM.
In these theories each SM field is associated to operators of the composite sector with well defined 
quantum numbers under the global symmetry $G$ of the composite sector. Most minimally $G$ contains
$SU(2)_L\otimes SU(2)_R\otimes U(1)_X$ but this symmetry will be enlarged in the most compelling scenarios
where the Higgs is a Goldstone boson. For simplicity  we will frame our discussion for a Goldstone boson Higgs in what follows.

In the minimal scenario each SM chirality couples to a single operator of the composite sector with equal quantum numbers 
under the SM gauge symmetries. For the quark sector on which we focus we have,
\begin{equation}
{\cal L}_{mix}=\lambda_L^{ij} \bar{q}_L^i Q_R^j+\lambda_{Ru}^{ij} \bar{U}_L^i  u_R^j+\lambda_{Rd}^{ij} \bar{D}_L^i  d_R^j 
\end{equation}
where lower-case letters refer to SM fields and the capital letters are fragments of multiplets of the strong sector. 
For the composite states we will write the most general lagrangian compatible with the symmetries.
In particular this contains couplings between different representations of the strong sector, $Y_A$, allowed by the spontaneous breaking of the Goldstone symmetry.
With abuse of language we will refer to these as the Yukawas of the strong sector. 
Upon mixing with the elementary fermions the SM Yukawas are generated,
\begin{eqnarray}
y_u&=&  \epsilon_L \cdot Y_u \cdot \epsilon_{Ru}\nonumber \\
y_d&=& \epsilon_L  \cdot Y_d\cdot   \epsilon_{Rd}
\end{eqnarray}
where we introduced the mixing matrices $\epsilon= \frac {\lambda} {g_\rho}$.

The maximal flavor symmetry that the  elementary+composite action can enjoy is,
\begin{equation}
U(3)_{el}^3\otimes U(3)_{comp}^3
\end{equation}
If unbroken this symmetry forbids both strong sector Higgs couplings and elementary-composite mixings.
The couplings that break these symmetries are the elementary-composite mixings, the strong sector Yukawas 
and non-degenerate mass terms of the strong sector. The couplings transform under the global symmetries above as,
\begin{eqnarray}
\epsilon_L=(3,1,1,\bar{3},1,1)\nonumber \\
\epsilon_{Rd}=(1,\bar{3},1,1,3,1)\nonumber \\
\epsilon_{Ru}=(1,1,\bar{3},1,1,3)\nonumber \\
Y_d=(1,1,1,3,\bar{3},1)\nonumber\\
Y_u=(1,1,1,3,1,\bar{3})
\end{eqnarray}
One can perform the counting of physical flavor structures similarly to the SM.
In general there are 5 complex matrices which can be diagonalized by means of 10 unitary matrices.
With a field redefinition of elementary and composite states we can eliminate 6 of these matrices. 
It follows that in general there are 4 physical unitary matrices that violate flavor in anarchic composite Higgs models. 
In the scenarios with anarchic strong sector considered in the literature the light generations have small
compositeness and this suppresses the dangerous flavor transitions. Despite this mechanism some tension
exists in several observables, see \cite{compositeMFV} for a summary. 

We will also be interested in more general setups where the SM quarks couple to several states
of the composite sector. In particular an interesting scenario is the one where the right handed quarks couple 
to singlets of custodial symmetry. In this case the unbroken $U(1)_X$ symmetry of the composite sector requires that
the left-handed quarks  couple to two different representations in order to generate masses for the up and down quarks,
\begin{eqnarray}
y_u&=&  \epsilon_{Lu} \cdot Y_u \cdot \epsilon_{Ru}\nonumber \\
y_d&=& \epsilon_{Ld}  \cdot Y_d\cdot   \epsilon_{Rd}
\end{eqnarray}
The maximal symmetry of the elementary + composite lagrangian is now
\begin{equation}
U(3)_{el}^3\otimes U(3)_{comp}^4
\end{equation}
with the couplings transforming as,
\begin{eqnarray}
\epsilon_{Lu}=(3,1,1,\bar{3},1,1,1)\nonumber \\
\epsilon_{Ld}=(3,1,1,1,\bar{3},1,1)\nonumber \\
\epsilon_{Rd}=(1,\bar{3},1,1,1,3,1)\nonumber \\
\epsilon_{Ru}=(1,1,\bar{3},1,1,1,3)\nonumber \\
Y_d=(1,1,1,1,3,\bar{3},1)\nonumber\\
Y_u=(1,1,1,3,1,1,\bar{3})
\end{eqnarray}

One can repeat the counting of flavor structures to find that there is an extra unitary matrix that violates flavor. 
This corresponds to the relative matrix that diagonalizes $\epsilon_{Lu}$ and $\epsilon_{Ld}$ on the left.
We emphasize that this matrix may jeopardize the suppression of flavor transitions associated to partial compositeness \cite{Csaki:2008eh}. 
Indeed recall that a crucial assumption for the mechanism to work is that the rotation matrices to the quark mass basis must be hierarchical, in particular the left rotations should have similar hierarchies to the CKM matrix. 
This is automatic when $q_L$ couples to a single representation. If instead $q_L$ couples to two representations some alignment between the two left mixings must be required to avoid large contributions from the operators associated to the third generation before rotation to the physical basis. For this reason some popular models in the literature, such us the Minimal Composite Higgs with fermions in the 5 of $SO(5)$ \cite{custodian}, suffer in general of a flavor problem.

\section{Realizations of Composite MFV}
\label{compositeMFV}

A very different realization of flavor was proposed in \cite{compositeMFV} (see also \cite{Cacciapaglia:2007fw,duccio,rattazzizaffaroni,perez} 
for related work) where the framework of partial compositeness was used to realize MFV.
Here all the flavor violation originates from the SM Yukawas, i.e.  a single flavor structure  exists, 
the CKM matrix. This can be realized in various ways:

\subsection{Partially Composite Quarks}

One possibility is that the strong sector respects a flavor symmetry that forbids flavor transitions \cite{compositeMFV}.
In the most minimal scenario one considers an $U(3)_F$ invariant strong sector so that $Y_{u,d}$ can be chosen proportional to the identity. 
There are 3 spurion matrices ($\epsilon_L\,,   \epsilon_{Ru} \,, \epsilon_{Rd}$)  that break the flavor symmetries.  
Under the SM + strong sector flavor symmetry they transform as, 
\begin{eqnarray}
\epsilon_{L}=(3,1,1,\bar{3})\nonumber \\
\epsilon_{Rd} =(1,\bar{3},1,3) \nonumber \\
\epsilon_{Ru}= (1,1, \bar{3}, 3)
\end{eqnarray}
MFV can be realized if $\epsilon_L$ respects the flavor symmetry, $\epsilon_L \propto Id$ in the basis above.
In this case $\epsilon_{Ru}$ and $\epsilon_{Rd}$ are proportional to the SM Yukawas.
This model suffers strong constraints from precisions tests \cite{compositeMFV}. In particular since compositeness of the 
left-handed up quarks is determined by the one of the top, from the hadronic width of the $Z$ measured
at per mille level at LEP one derives,
\begin{equation}
\epsilon_L\lsim \frac {1} {g_\rho} \left(\frac {m_\rho}{6\, {\rm TeV}}\right)
\label{leftbound}
\end{equation}
and a similar bound follows from unitarity of the CKM matrix. This puts  significant pressure on the model 
in order to reproduce the required top mass.

The alternative option to realize MFV is that the SM flavor structure is carried by the left mixings.
This requires that the composite sector respects a flavor symmetry $U(3)_U\otimes U(3)_D$
that can be realized when the right quarks couple to singlets of the strong sector, see \cite{compositeMFV} for the details.
The mixings now transform as,
\begin{eqnarray}
\epsilon_{Lu}=(3,1,1,1,\bar{3})\nonumber \\
\epsilon_{Ld}=(3,1,1,\bar{3},1)\nonumber \\
\epsilon_{Rd} =(1,\bar{3},1,3,1) \nonumber \\
\epsilon_{Ru}= (1,1, \bar{3},1, 3)\,.
\end{eqnarray}
When the right-handed quarks couple to singlets there are no tree level corrections to their couplings,
allowing the model to be compatible with electro-weak precision tests. In this case the main bounds 
come from searches of compositeness at the LHC, in particular di-jets \cite{compositeMFV,Domenech:2012ai}.
Using the latest limits \cite{CMS2.2comp} we estimate,
\begin{equation}
\epsilon_{Ru}^2 \lsim    \frac {1.5\,\lambda_t} {g_\rho} \left(\frac {m_{\rho}}{3~{\rm TeV}} \right)
\label{compbound}
\end{equation}
and a slightly weaker bound applies to down quarks. Large compositeness is excluded if the coupling is maximal.
A lower bound $\epsilon_{Ru}\gsim \lambda_t/ g_\rho$ must be satisfied to reproduce the top mass.

\subsection{Fully Composite Quarks and other Models}

Another realization of MFV can be obtained when all the quarks are composite. This was proposed long ago
by Rattazzi and Zaffaroni (RZ) \cite{rattazzizaffaroni}. If all the quarks are composite then there are only 
the composite sector flavor symmetries. One requires the flavor symmetry to be as in the SM,
\begin{equation}
U(3)_{comp}^3
\end{equation}
broken by two bi-fundamentals fields that acquire a VEV.  If no other sources of flavor violation exist this realizes MFV. 
Unfortunately this model suffers from very severe bounds from precision tests. 
In particular from the coupling of left quarks to the $Z$ we can estimate the required compositeness scale,
\begin{equation}
m_\rho \gsim 6\, g_\rho\, {\rm TeV}
\end{equation}
Also direct bounds from compositeness are rather severe in this case and incompatible with  the scale demanded by naturalness.

The discussion in the previous section suggests a variant of the RZ model with partially composite quarks. 
To realize this the elementary-composite mixings must respect the diagonal elementary-composite flavor symmetry. 
Since the left-handed light quarks have compositeness related to the one of the top, this model 
will at best have bounds similar to the scenario with left-handed compositeness in \cite{compositeMFV}. 

To complete the possibilities one could also construct a hybrid RZ model with,
\begin{equation}
Y_u \propto y_u\,,~~~~~~~~~~~~~~~~~~~ \epsilon_{Rd} \propto y_d
\end{equation}
and all the other couplings flavor trivial. This choice has similar constraints as left-handed compositeness.
Extending the symmetry of the strong sector to $U(3)^4_c$  we could 
for example place the Yukawa structure in the up left mixings and  $Y_d$. 
In this case bounds are similar to scenario with right-handed compositeness. In general it does not seem
possible to get weaker bounds than in right-handed compositeness if the theory realizes MFV.

\subsection{Power Counting}

The MFV hypothesis guarantees that the minimum amount of flavor violation beyond the SM is generated
but there can still be important flavor constraints if the scale of new physics is in the TeV range 
and effective operators are generated at tree level. In our realizations of MFV a specific pattern 
of effective operators, determined by the scale $m_\rho$ and coupling $g_\rho$ of the strong sector is obtained.
We report the results in Table \ref{table:MFVbounds}, for some of the most constrained MFV effective operators.
In particular for the case of right-handed compositeness we find that flavor bounds from $K-\bar{K}$ mixing 
are relevant.

\begin{table}[t]
\begin{center}
\begin{tabular}{c||c|c|c|c}
 {\rm operator}   & R-MFV\,  & L-MFV \, & RZ-MFV & $\Lambda ({\rm TeV}) $\,  \\  \hline \hline
$H^\dagger (\bar{d}_R y_d^\dagger y_u y_u^\dagger \sigma_{\mu\nu}q_L) e F^{\mu\nu}$ 		 & $\sim 0$ 	& $\sim 0$ & $4\pi\, m_\rho$  & 6.1 \\   
$\frac 1 2 (\bar{q}_L y_u y_u^\dagger \gamma_\mu q_L)^2$ 		 & $g_\rho \, m_\rho\, \epsilon_{Ru}^2$ & $\sim 0$  & $g_\rho \, m_\rho$ & 5.9 \\   
$i(q_L y_u y_u^\dagger \gamma_\mu q_L) H^\dagger D_\mu H$ 		 & $m_\rho\, \epsilon_{Ru}$ & $\sim 0$ & $m_\rho$ & 2.3\\   
$i(q_L y_u y_u^\dagger \gamma_\nu q_L) e D_\mu F^{\mu\nu}$ 		 &  $g_\rho\,m_\rho\, \epsilon_{Ru}$ & $\sim 0$ &  $g_\rho\,m_\rho$ & 1.5 \\   
\end{tabular}
\end{center}
\caption{Suppression scale of MFV operators and experimental bound \cite{MFVbounds} in models with composite 
right-handed quarks (R-MFV), composite left-handed quarks (L-MFV) and Rattazzi-Zaffaroni model (RZ-MFV).}
\label{table:MFVbounds}
\end{table}

The power counting of effective operators can be understood in general adapting the results 
of Ref. \cite{silh} to flavor symmetries. Let us consider the case of fully composite quarks first. 
We will treat the Yukawas as expectation values of fields that transform as bi-fundamentals of the
flavor symmetry. The effective action takes the form \cite{silh},
\begin{equation}
{\cal L} =\frac {m_\rho^4}{g_\rho^2}\left[{\cal L}^0(\Phi, \frac {\partial}{m_\rho}, \frac {A}{m_\rho})+\frac {g_\rho^2}{16\pi^2}{\cal L}^1(\Phi, \frac {\partial}{m_\rho}, \frac {A}{m_\rho})+\dots \right]
\label{silh}
\end{equation}
where $\Phi$ are dimensionless fields (fermionic and bosonic), $A$ gauge fields and the expansions of ${\cal L}^{0,1,...}$ have order one coefficients.
In terms of canonically normalized fields, the Yukawas of the strong sector are given by,
\begin {equation} 
y_{SM}= g_\rho^2 \frac {\langle Y\rangle}{m_\rho}
\label{yRZ}
\end{equation}
Let us now consider a general MFV effective operator,
\begin{equation}
\psi^n \partial^m  H^k  y_{SM}^l
\end{equation}
From the general structure (\ref{silh}) and using (\ref{yRZ}) we find that the coefficient scales as,
\begin{equation}
\frac {g_\rho^{n+k-l-2}}{m_\rho^{m+k+\frac  3 2 n-4}}
\end{equation}
times a loop factor if the operator is generated at loop level. For example we will assume that dipole operators are generated
at 1-loop. As an example consider effective flavor preserving 4-fermi operators such as,
\begin{equation}
(\bar{q}  \gamma_\mu q)^2
\end{equation}
We find that the coefficient scales as,
\begin{equation}
\frac {g_\rho^2}{m_\rho^2}
\end{equation}
The operator above is for example generated through the exchange of heavy vector fields.
Powers of the SM Yukawas come with inverse powers of $g_\rho$ so that the MFV operator,
\begin{equation}
(\bar{q}_L y_u y_u^\dagger  \gamma_\mu q_L)^2
\label{LLMFV}
\end{equation}
will be suppressed by
\begin{equation}
\frac 1 {g_{\rho}^2 m_\rho^2}
\end{equation}
and higher order terms in the MFV expansion will be suppressed by higher powers of $g_\rho$.

This analysis can be extended to the case of MFV with partially composite quarks. 
The SM effective operators are now obtained by dressing the strong sector (flavor diagonal) ones with the mixings.
Consider for example right-handed compositeness. In this case we find that the coefficient of 
the effective operator  (\ref{LLMFV}) is given by,
\begin{equation}
\frac 1 { g_\rho^2 m_\rho^2 \epsilon_{Ru}^4}
\end{equation}
so it is enhanced with respect to the case of fully composite quarks. From the experimental bound we derive,
\begin{equation}
\epsilon_{Ru}^2> \frac 2 {g_\rho}\left(\frac{3\, TeV}{m_\rho}\right).
\end{equation}
This  scales in opposite way with respect compositeness bounds (\ref{compbound}) and
could be an important constraint. The same conclusion applied in the RZ model with partially composite quarks.
Of course these are just rough estimates and important numerical factor will be model dependent.

Note that extra selection rules exist in partial compositeness. 
If the Yukawa structure is carried either by the couplings of the left-handed or right-handed quarks, insertions of mixings and Higgs 
are required to generate certain operators. For this reason the flavor bounds on left-handed compositeness are extremely mild. 
Indeed one can see that as in the SM there no tree level flavor changing neutral currents \cite{Cacciapaglia:2007fw,pisa}. 
These only arise at loop level and they are further suppressed by mixings.  On the contrary for right-handed compositeness MFV
operators such as (\ref{LLMFV}) are generated at tree level leading to important constraints. Therefore  left-handed compositeness 
is the safest  scenario for flavor but is disfavored by precision tests while right-handed compositeness is the opposite. 
We estimate the size of some of the most relevant MFV operators in table \ref{table:MFVbounds}.

We expect very mild bounds on operators involving leptons. For example,
\begin{equation}
(\bar{q}_L y_u y_u^\dagger \gamma^\mu q_L) (\bar{e}_R \gamma_\mu e_R)
\end{equation}
must be suppressed by a scale of at least 2.7 TeV \cite{MFVbounds}. 
This is easily achieved if the leptons have compositeness smaller than quarks.

\section{Beyond MFV}
\label{beyondMFV}

The above realizations of MFV are already constrained by the data. Right-handed compositeness is 
at present the most attractive scenario but experimental limits will improve 
with LHC measurements already in 2012 where a significant fraction of parameters space will be tested. 
This is possible because, contrary to the anarchic scenario, MFV requires compositeness of right-handed quarks
 to be large, being related  to the one of the top by the flavor  symmetry. 
As a consequence direct searches, in particular di-jets are particularly relevant for
this scenario \cite{compositeMFV,boundsr}.

While MFV is the flavor safest world, other variants could be sufficient phenomenologically
\footnote{In the context of supersymmetry it was shown that the symmetry $U(2)^3$ 
acting of the first two generation is sufficient  phenomenologically if
the SM flavor structure is  generated by the following spurions \cite{barbieri}:
\begin{equation}
\Delta Y_u= (2,\bar{2},1)\,,~~~~~ \Delta Y_d=(2,1,\bar{2})~~~~\lambda=(2,1,1)
\end{equation}
Our construction will be different because we will write the most general elementary-composite coupling allowed by 
the symmetries but we will require a smaller flavor symmetry of the strong sector than in \cite{compositeMFV,rattazzizaffaroni}.}.
In order to avoid the constraints it is necessary to deviate from exact MFV treating the third 
generation differently. This would indeed be natural if one chirality of the top quark is part of the strong sector as in this case 
the elementary flavor symmetry will be reduced $U(3)^2\otimes U(2)$ but we will consider this possibility in general.

\subsection{Right-handed Compositeness}

The first model that we consider is based on the subgroup of the SM flavor symmetries,
\begin{equation}
U(3)_L\otimes  U(3)_{Rd}\otimes U(2)_{Ru}
\end{equation}
which could be motivated by full $t_R$ compositeness.

To realize this pattern we assume that the strong sector has a symmetry,
\begin{equation}
U(3)_{D}\otimes U(2)_{U}
\end{equation}
under which the partners of the top are a singlets, the partners of up and charm are doublets of $U(2)_{U}$ and the partners of the 
down quarks are triplets of $U(3)_D$. The strong sector has no flavor violating structures, however $Y_u$ and the masses of the up 
sector partners do not need to be universal. If the mixings of the right-handed quarks respect this symmetry, the elementary+composite lagrangian will 
have a symmetry $U(3)_L\otimes U(3)_{Rd}\otimes U(2)_{Ru}$ only broken by the left mixings. 
Note that, as in right-handed MFV, this choice requires  that the elementary left doublet couples to two states. We can choose the basis,
\begin{eqnarray}
\epsilon_{Lu}&=&\hat{\epsilon}_{Lu}\cdot U_1 \nonumber \\
\epsilon_{Ld}&=&U_2\cdot \hat{\epsilon}_{Ld}\nonumber \\
\epsilon_{Ru}&=&\hat{\epsilon}_{Ru} \nonumber \\
\epsilon_{Rd}&=&\hat{\epsilon}_{Rd}\nonumber \\
Y_u&=&\hat{Y}_u\nonumber \\
Y_d&=&\hat{Y}_d\,.
\end{eqnarray}
where $\hat{\epsilon}$ are diagonal matrices. $\hat{\epsilon}_{Rd}$ is proportional to the identity by the flavor symmetry so 
 $\hat{\epsilon}_{Ld}$ is proportional to the down quark masses.  In the up sector the matrix $\hat{\epsilon}_{Ru}$ is not universal,
\begin{equation}
\hat{\epsilon}_{Ru} =  {\rm Diag}[a,a,b]
\end{equation} 
The same is true for $\hat{Y}_u$ but for simplicity we assume it to be flavor universal ($=g_\rho$) in our estimates.
This corresponds to a strong sector $U(3)_U$ invariant with the symmetry broken to $U(2)_U$ by the up right mixing. 
The hierarchies of $\hat{\epsilon}_{Lu}$ are then determined on average by the masses of the up quarks.
Note that this also determines the  hierarchies of the left rotation matrix in the up sector.
This is typically very close to the identity (more than the CKM matrix). 
As a consequence the matrix $U_2$ is essentially the CKM matrix.
Importantly the rotation matrix $D_R$ of the down right quarks to the mass basis is the identity. 
For this reason the contribution to  dangerous LR 4-Fermi operators in the down sector is suppressed as in MFV.

We consider two examples R$_1$ and R$_2$ corresponding to the following
choices of parameters, 
\begin{center}
R$_1$:~~~~~~
\begin{tabular}{|c|c|c|c|c|c|c|c|c}
\hline
$m_\rho$~(TeV) & $g_{\rho}$ & $\hat{\epsilon}_{Ru}$ & $\hat{\epsilon}_{Rd}$ \\ 
\hline 
$3$ & $5$ & $(0.1,0.1, 0.5)$ & $(0.1,0.1, 0.1)$ \\ \hline
\end{tabular}
\end{center}
\begin{center}
R$_2$:~~~~~~~
\begin{tabular}{|c|c|c|c|c|c|c|c|c}
\hline
$m_\rho$~(TeV) & $g_{\rho}$ & $\hat{\epsilon}_{Ru}$ & $\hat{\epsilon}_{Rd}$ \\ 
\hline 
$3$ & $5$ & $(0.05,0.05, 0.5)$ & $(0.05,0.05, 0.05)$ \\ \hline
\end{tabular}
\end{center}
where to obtain our estimates we take the strong sector Yukawas equal $g_\rho$.
We will also report the MFV scenario where all the right mixings are 0.5.

We consider the following 4-Fermi operators,
\begin{equation}
\begin{aligned}
Q_1^{q_i q_j}&={\overline q}^\alpha_{jL} \gamma_\mu q_{iL}^\alpha\,{\overline q}^\beta_{jL} \gamma^\mu q_{iL}^\beta \,,\\
\tilde Q_1^{q_i q_j}&={\overline q}^\alpha_{jR} \gamma_\mu q_{iR}^\alpha\,{\overline q}^\beta_{jR} \gamma^\mu q_{iR}^\beta \,,\\
Q_4^{q_i q_j}&={\overline q}^\alpha_{jR} q_{iL}^\alpha\,{\overline q}^\beta_{jL} q_{iR}^\beta \,.
\end{aligned}
\end{equation}
that are obtained from the structures analog to (\ref{LLMFV}) (generated for example by heavy vector boson exchange). 
Parametrizing the effective lagrangian as $\sum_i C_i\, O_i$ we obtain the following estimates, obtained scanning over parameters that reproduce the SM Yukawas, for the Wilson coefficients of $\Delta F=2$ operators,
\begin{table}[th]
\begin{center}
\begin{tabular}{c||c|c|c|c}
 (in GeV$^{-2}$)  & R$_1$\,& R$_2$\,  & R-MFV \, & EXP \\  \hline
$C_4^K$ 		& 0 & 0	& 0 & $(700+4i)\cdot 10^{-17}$ \\ 
$C_4^D$ 	& $(5+5 i) \cdot 10^{-16}$	& $ (5+5 i) \cdot 10^{-16}$	& 0 & $8(1+i) \cdot 10^{-14}$ \\ 
$C_4^{B_d}$  & 0 & 0	& 0 & $3(1+i)\cdot 10^{-13}$ \\
$C_4^{B_s}$ 	& 0	 & 0	& 0 & $2(1+i)\cdot 10^{-11}$ \\ \hline
$C_1^K$ 	& $(4+4 i)\cdot 10^{-15}$ 	  & $ (10+5 i) \cdot 10^{-15}$    &       $(2+2 i) \cdot 10^{-15}$ & $(1000+4 i)\cdot 10^{-15}$ \\
$C_1^D$ 	& 0	&   0	 & 0 & $(7+7i) \cdot 10^{-13}$ \\
$C_1^{B_d}$  & $(1+i) \cdot 10^{-12}$  & $ (2+2 i) \cdot 10^{-12}$ & $(1+ i)\cdot 10^{-12}$ & $(1+2i)\cdot 10^{-11}$ \\
$C_1^{B_s}$ 	& $(4+ i) \cdot 10^{-11}$ &$ (4+2 i) \cdot 10^{-11}$	& $(40+i)\cdot 10^{-12}$ & $ (1+i)\cdot 10^{-9}$ \\  \hline
$\tilde{C}_1^K$ 	&  0 	           & 0 &       0 & $(1000+4 i)\cdot 10^{-15}$ \\
$\tilde{C}_1^D$ 	&   $(1+ i) \cdot 10^{-11}$	& $(5+5 i) \cdot 10^{-13}$	& 0 &$(7+7i)\cdot 10^{-13}$ \\
$\tilde{C}_1^{B_d}$  & 0 &0 &  0 & $(2+2i)\cdot 10^{-11}$ \\
$\tilde{C}_1^{B_s}$ 	& 0 & 0	& 0 &  $ (1+i)\cdot 10^{-9}$ \\  \hline
\end{tabular}
\end{center}
\caption{Average values of Wilson coefficients of $\Delta F=2$ operators in models with right-handed compositeness and their experimental limit \cite{utfit}.
Values smaller than $10^{-19}$ GeV$^{-2}$ are rounded to zero.}
\label{table:wilsonF2ur}
\end{table}\\
As expected this realization of flavor differs from MFV in the up sector. Indeed depending on the choice of parameters some tension exists with $\tilde{C}_1^D$ (the RR operator contributing to $D-\bar{D}$ mixing). $C_1^K$ is always close to the bound and give the main flavor bound in MFV.

We have also considered $\Delta F=1$ dipole operators contributing in particular to the $b\to s \gamma$ transition.
The largest contribution arises from the first operator as in Table \ref{table:MFVbounds} and has similar size as in MFV so 
it does not provide a constraint on the model. The contributions from penguin diagrams which are sizable within anarchic scenarios
are negligible.

\newpage

\subsection{Left-handed Compositeness}

The second model that we consider is based on the subgroup of the SM flavor symmetries,
\begin{equation}
U(2)_L\otimes  U(3)_{Rd}\otimes U(3)_{Ru}
\label{symmetryleft}
\end{equation}
which would be natural if $t_L$ were fully composite.  To realize this pattern we consider a strong sector with a symmetry,
\begin{equation}
U(2)_F
\end{equation}
under which the composite states transform as doublets (partners of first two generations) 
and singlets (partners of the third generation). The flavor symmetry (\ref{symmetryleft}) is realized 
when the mixing with the elementary left doublets respects $U(2)_F$,
\begin{equation}
\hat{\epsilon}_{L} =  {\rm Diag}[a,a,b]
\end{equation}
This generalizes the MFV model with left-handed compositeness. 
Upon field redefinitions we can choose the mixings as,
\begin{eqnarray}
\epsilon_{Lu}&=&\hat{\epsilon}_{L} \nonumber \\
\epsilon_{Ru}&=&U_1\cdot \hat{\epsilon}_{Ru} \nonumber \\
\epsilon_{Rd}&=&U_2 \cdot \hat{\epsilon}_{Rd}\nonumber \\
Y_u&=&\hat{Y}_u\nonumber \\
Y_d&=&\hat{Y}_d
\end{eqnarray}
To reproduce the CKM matrix, denoting with $U_L$  the matrix that diagonalizes the up Yukawas,  
we have $U_2\approx U_L\cdot V_{CKM}$. Moreover, due to the hierarchies of quark masses, the rotation matrices of
the right-handed quarks are very close to the identity suppressing contributions to LR operators.

We consider two representative examples,
\begin{center}
L$_1$:~~~~~~
\begin{tabular}{|c|c|c|c|c|c|c|c}
\hline
$m_\rho$~(TeV) & $g_{\rho}$ & $\hat{\epsilon}_{L}$  \\ 
\hline 
$3$ & $5$ & $(0.1,0.1, 0.5)$  \\ \hline
\end{tabular}
\end{center}
\begin{center}
L$_2$:~~~~~~
\begin{tabular}{|c|c|c|c|c|c|c|c}
\hline
$m_\rho$~(TeV) & $g_{\rho}$ & $\hat{\epsilon}_{L}$ \\ 
\hline 
$3$ & $5$ & $(0.05,0.05, 0.5)$  \\ \hline
\end{tabular}
\end{center}

We present our estimates of  $\Delta F=2$ operators in Table \ref{table:wilsonF2ul}.
All the bounds are easily satisfied.  $\Delta F=1$ transitions are also suppressed as in right-handed compositeness.
The constraint from precision test on the left mixing (\ref{leftbound}) applies for the light quarks 
that therefore cannot be strongly composite but contrary to the MFV scenario  this does not lead to a tension to generate the top mass.

\begin{table}[th]
\begin{center}
\begin{tabular}{c||c|c|c|c}
 (in GeV$^{-2}$)  & L$_1$\,& L$_2$\,  & L-MFV \, & EXP \\  \hline
$C_4^K$ 		& 0 &0	& 0 & $(700+4)\cdot 10^{-17}$ \\ 
$C_4^D$ 	& 0	&  0	& 0 & $8(1+i) \cdot 10^{-14}$ \\ 
$C_4^{B_d}$  & $(3+ 3 i) \cdot 10^{-18}$ & $(3+ 3 i) \cdot 10^{-18}$		& 0 & $3(1+i)\cdot 10^{-13}$ \\
$C_4^{B_s}$ 	& $(9+ 9 i) \cdot 10^{-16}$	 &  $(7+ 7 i) \cdot 10^{-16}$	& 0 & $2(1+i)\cdot 10^{-11}$ \\ \hline
$C_1^K$ 	& $(30+5 i)\cdot 10^{-17}$ 	  &    $(10+5 i)\cdot 10^{-18}$        &       0 & $(1000+4 i)\cdot 10^{-15}$ \\
$C_1^D$ 	& $(5+5 i)\cdot 10^{-15}$	& $(5+5 i)\cdot 10^{-15}$	& 0 & $(7+7i)\cdot 10^{-13}$ \\
$C_1^{B_d}$  & $(8+8 i) \cdot 10^{-13}$  & $(2+2 i) \cdot 10^{-13}$ & 0 & $(2+2i)\cdot 10^{-11}$ \\
$C_1^{B_s}$ 	& $(1+ i) \cdot 10^{-11}$ &  $(3+ 3 i) \cdot 10^{-12}$	& 0 & $(1+i)\cdot 10^{-9}$ \\  \hline
$\tilde{C}_1^K$ 	& 0	           & 0 &       0 & $(1000+4 i)\cdot 10^{-15}$ \\
$\tilde{C}_1^D$ 	&   0 	&  0 & 0  &$(7+7 i)\cdot 10^{-13}$ \\
$\tilde{C}_1^{B_d}$  & 0  & 0 & 0 & $(2+2 i)\cdot 10^{-11}$ \\
$\tilde{C}_1^{B_s}$ 	& 0  & $0$ & 0   &  $ (1+i)\cdot 10^{-9}$ \\  \hline
\end{tabular}
\end{center}
\caption{Average values of Wilson coefficients of $\Delta F=2$ operators in models with left-handed compositenss
and their experimental limit \cite{utfit}. Values equal or smaller than $10^{-19}$ GeV$^{-2}$ are rounded to zero.}
\label{table:wilsonF2ul}
\end{table}

\newpage

\subsection{Rattazzi-Zaffaroni Scenarios}

We can also realize an extension of RZ model based on a similar pattern of symmetries. 
To do this we assume that the mixings that respect the flavor symmetry are,
\begin{equation}
\epsilon_i =  {\rm Diag}[a_i,a_i,b_i]
\end{equation}
Various possibilities exist depending on the flavor symmetry of the strong sector. We will consider two representative 
choices. For the first one similar to right-handed compositeness we choose,
\begin{center}
$U(3)_L\otimes U(2)_{Rd} \otimes U(2)_{Ru}$:~~~~~~
\begin{tabular}{|c|c|c|c|c|c|c|c|c|c}
\hline
$m_\rho$~(TeV) & $g_{\rho}$ & $\hat{\epsilon}_{L}$ & $\hat{\epsilon}_{Ru}$ & $\hat{\epsilon}_{Rd}$\\ 
\hline 
$3$ & $5$ & $(0.5,0.5, 0.5)$ & (0.05,0.05, 0.5) & (0.05,0.05, 0.5)\\ \hline
\end{tabular}
\end{center}
and for the second similar to left-handed compositeness we take,
\begin{center}
$U(2)_L \otimes U(3)_{Rd} \otimes U(3)_{Ru}$:~~~~~~
\begin{tabular}{|c|c|c|c|c|c|c|c|c|c}
\hline
$m_\rho$~(TeV) & $g_{\rho}$ & $\hat{\epsilon}_{L}$ & $\hat{\epsilon}_{Ru}$ & $\hat{\epsilon}_{Rd}$\\ 
\hline 
$3$ & $5$ & $(0.1,0.1, 0.5)$ & (0.5,0.5, 0.5) & (0.5,0.5, 0.5)\\ \hline
\end{tabular}
\end{center}
For comparison we also consider the MFV scenario with all the mixings taken 0.5.

The estimates of  $\Delta F=2$ operators are reported in the table \ref{table:wilsonF2RZ}.
The first model, similarly to models with right-handed compositeness, gives sizable contributions to the  RR operators contributing 
to the $K$,  $D$  and $B$ mixings that are close to the experimental bound for this choice of parameters. 
We have also considered the  model with $U(2)_L\otimes U(2)_{Rd}\otimes U(2)_{Ru}$ finding that it suffers from 
a severe flavor problem.
\begin{table}[th]
\begin{center}
\begin{tabular}{c||c|c|c|c}
 (in GeV$^{-2}$)  & $U(3)_L\otimes U(2)^2_R$ \,&  \,  $U(2)_L\otimes U(3)^2_R$ \, & RZ-MFV\,& EXP \\  \hline
$C_4^K$ 		& 0 & 0	&  0 & $(700+4)\cdot 10^{-17}$ \\ 
$C_4^D$ 	& 0	&  0	&  0 & $8(1+i) \cdot 10^{-14}$ \\ 
$C_4^{B_d}$  & 0 & $(4+4 i) \cdot 10^{-18}$		&  $(4+4 i) \cdot 10^{-18}$ & $3(1+i)\cdot 10^{-13}$ \\
$C_4^{B_s}$ 	& 0	 &  $(1+ i) \cdot 10^{-15}$ &  0 & $2(1+i)\cdot 10^{-11}$ \\ \hline
$C_1^K$ 	& $(10+5 i)\cdot 10^{-15}$ 	  &    $(4+2 i)\cdot 10^{-15}$        &       $(2+2 i) \cdot 10^{-15}$ & $(1000+4 i)\cdot 10^{-15}$ \\
$C_1^D$ 	& 0	& $(8+8 i)\cdot 10^{-15}$	& 0 & $(7+7 i)\cdot 10^{-13}$ \\
$C_1^{B_d}$  & $(1+ i) \cdot 10^{-12}$  & $ (2+2 i) \cdot 10^{-12}$ & $(1+ i)\cdot 10^{-12}$ & $(2+2i)\cdot 10^{-11}$ \\
$C_1^{B_s}$ 	& $(4 +2  i) \cdot 10^{-11}$ &  $(4+ 2 i) \cdot 10^{-11}$	& $(40+8 i)\cdot 10^{-12}$ & $(1+i)\cdot 10^{-9}$ \\  \hline
$\tilde{C}_1^K$ 	& $(5+7 i) \cdot 10^{-15}$	 & 0 &        0 & $(1000+4 i)\cdot 10^{-15}$ \\
$\tilde{C}_1^D$ 	&   $(3+3 i) \cdot 10^{-13}$ 	& 0 & 0  &$(7+7i)\cdot 10^{-13}$ \\
$\tilde{C}_1^{B_d}$  & $(5+5 i) \cdot 10^{-11}$  &  0 & 0 & $(2+2 i)\cdot 10^{-11}$ \\
$\tilde{C}_1^{B_s}$ 	& $(1+ i) \cdot 10^{-11}$  & 0 & 0   &  $(1+i)\cdot 10^{-9}$ \\  \hline
\end{tabular}
\end{center}
\caption{Average values of Wilson coefficients $\Delta F=2$ operators in RZ models and their experimental limit \cite{utfit}.
Values equal or smaller than $10^{-19}$ GeV$^{-2}$ are rounded to zero.}
\label{table:wilsonF2RZ}
\end{table}\\

\subsection{Phases}

The scenarios under consideration contain several new flavor diagonal CP violating phases.
Consider for example the scenario of left-handed compositeness, similar arguments can be repeated for the others. 
The flavor symmetry $U(2)_L\otimes U(3)_{Rd} \otimes U(3)_{Ru}$ is broken by the right mixings which are two 
general complex matrices. Divided into real a complex they contain (18,18) parameters. 
The flavor symmetry allows to remove $(7,15)$ parameters, corresponding to the number of (broken) generators. 
Therefore the number of physical parameters is,
\begin{equation}
(11,3)
\end{equation}
In particular there are at least 2 new physical CP phases beside the CKM phase.  In addition to these there could also be CP phases 
in the strong sector.

Interestingly to leading order these new sources of CP violation do not contribute to EDMs for the same reasons explained in \cite{compositeMFV}.
This should be compared with the situation in the anarchic scenario where these contribution are typically large and lead to a tension with EDMs. 
To see how this works the leading contribution to the EDM of the down  quark is given  by,
\begin{eqnarray}
d_d \sim   \displaystyle{   \frac 1 {32 \pi^2} \frac v {m_\rho^2}}\, {\rm Im}\left[D_L^\dagger \cdot \epsilon_L\cdot (a_1\, Y_d \cdot Y^{\dagger}_d+a_2\, Y_u \cdot Y^{\dagger}_u)\cdot Y_d \cdot \epsilon_{Rd} \cdot D_R\right]_{11}
\label{edmd}
\end{eqnarray}
where $a_{1,2}$ are order one model dependent coefficients. If $Y_{u,d}$ are proportional to the identity 
this is manifestly real being proportional to the SM Yukawas. Even in the more general case a very strong
suppression is obtained. Beside this contribution there can be also a UV dependent contribution to EDMs
generated by the strong sector. This would be zero if the strong sector respects CP.

\section{Conclusion}
\label{outlook}

The framework of partial compositeness allows new realizations of flavor in strongly coupled theories
that will be tested at the LHC. In the popular anarchic scenario light fermions are elementary which ameliorates
precision tests but some tension exists with flavor observables and EDMs. An alternative possibility is that MFV is realized,
in which case contributions to flavor observables are suppressed and show a specific structure determined by the coupling of the strong sector. 
MFV requires that the strong sector possesses flavor symmetries and that at least some  chiralities of light quarks 
are  strongly composite. Precision tests are more difficult to satisfy than in the anarchic scenario with the exception of right-handed compositeness. 
The phenomenology of composite MFV is very different and more visible at the LHC than anarchic scenarios, since production of resonances 
can be large due to the large compositeness of up and down quarks. Moreover compositeness bounds start to constrain significantly 
the most attractive scenario of right-handed compositeness.

In this note we have studied an intermediate scenario where we rely on a subgroup of the SM flavor symmetries deviating in this way from MFV.
This allows to treat the top quark differently, so that the light generations can be mostly elementary compatibly with the
flavor symmetries and would be particularly motivated if one chirality of the top is composite. 
This can be realized in a similar way to MFV if for example the strong sector has $U(2)$ flavor symmetries.
Flavor and EDMs bounds can be still satisfied very similarly to composite MFV. The scenario of left-handed compositeness that was
disfavored by precision tests is now viable and also compositeness bounds in right-handed compositeness are not severe.

We expect the phenomenology of these models to be rather similar to the one of the anarchic scenario. Since the light 
quarks are mostly elementary, production cross-sections of composite states will not be enhanced, so they will be tested only in the 14 TeV run of the 
LHC. The decays of the resonances will have a pattern determined by the flavor symmetries that might be interesting to investigate.
We leave a detailed study to future work.

\vspace{0.5cm} {\bf Acknowledgments:} 
I am grateful to Riccardo Rattazzi and Andreas Weiler for numerous discussion about flavor in composite Higgs models
and to Dario Buttazzo, Filippo Sala and David Straub for comments.
I would like to thank the GRAAL Film Lab in Athens for hospitality during part of this work.

\end{document}